\documentclass[aps,showpacs,twocolumn,floatfix,amsmath,preprintnumbers,nofootinbibe,secnumarabic,nopacs]{revtex4} 

\usepackage{graphicx,verbatim,amssymb,amsfonts,amsmath,amstext,hyperref,url}

\def\keyword#1{} \def\conflictsofinterest#1{} \def\funding#1{} \def\reftitle#1{} \def\boo#1#2#3#4#5#6{#1. \textit{#2}; #3: #4, #5.} \def\jrn#1#2#3#4#5#6{#1. #2. \textit{#3} \textbf{#6}, \textit{#4}, #5.} \def\andd{.; } \def\andt{.; } \def\Ref{} \def\Refs{}    \def\eqref#1{(\ref{#1})}   \def\scn#1#2{\section{#1}\lb{#2}}  \def\textendash{\hbox{-}} 

\def\boldsymbol#1{#1}

\def\bfl{\begin{flushleft}}
\def\efl{\end{flushleft}}
\def\bfr{\begin{flushright}}
\def\efr{\end{flushright}}
\def\bc{\begin{center}}
\def\ec{\end{center}}
\def\be{\begin{equation}}
\def\ee{\end{equation}}
\def\bse{\begin{subequations}}
\def\ese{\end{subequations}}
\def\ba{\begin{eqnarray}}
\def\ea{\end{eqnarray}}
\def\baa#1{\begin{array}{#1}}
\def\eaa{\end{array}}
\def\bw{\begin{widetext}}
\def\ew{\end{widetext}}

\def\lb#1{\label{#1}}
\def\bit{\begin{itemize}}
\def\eit{\end{itemize}}
\def\bco{}
\def\bcs{\begin{cases}}
\def\ecs{\end{cases}}

\def\schrod{Schr\"odinger}

\def\Der#1#2{\frac{\drm #1}{\drm #2}}
\def\pDer#1#2{\frac{\partial #1}{\partial #2}}

\def\vena{\boldsymbol{\nabla}}

\def\nc0{\tilde b_0}

\def\cf{{\cal D}}

\def\area{{\cal A}}
\def\vol{{\cal V}}
\def\drm{d}
\def\dvol{\drm\vol}

\def\text#1{{\rm #1}}

\begin{document}

\preprint{\small \footnotesize Fluids  \textbf{7}, 358 (2022)   
\ \ 
[\href{https://doi.org/10.3390/fluids7110358}{DOI: 10.3390/fluids7110358}]
}

\title{
Sound propagation in cigar-shaped Bose liquids in the Thomas-Fermi approximation: A comparative study between Gross-Pitaevskii and logarithmic models
}

\author{Konstantin G. Zloshchastiev}
\email{https://orcid.org/0000-0002-9960-2874} 
\affiliation{Institute of Systems Science, Durban University of Technology, P.O. Box 1334, Durban 4000, South Africa;
kostiantynz@dut.ac.za, kostya@u.nus.edu}

\begin{abstract} 
\noindent\textbf{Abstract}:
A comparative study is done of the propagation of sound pulses in elongated Bose liquids and Bose-Einstein condensates in Gross-Pitaevskii and logarithmic models, by means of the Thomas-Fermi approximation.
It is shown that in the linear regime the propagation of small density fluctuations 
is essentially one-dimensional in both models, in the direction perpendicular to the cross section of a liquid's lump.
Under these approximations, it is shown that the speed of sound scales as a square root of particle density in the case
of the Gross-Pitaevskii liquid/condensate, but it is constant in a case of the homogeneous logarithmic liquid.
\end{abstract}

\date{received: 30 September 2022}


\keyword{quantum fluid; Bose-Einstein condensate; logarithmic \schrod~equation; cigar-shaped Bose-Einstein condensate; Thomas-Fermi approximation; Gross-Pitaevskii equation}

\maketitle

\scn{Introduction}{s:in}
The remarkable experiments with cooling magnetically trapped alkali atoms 
down to nanokelvin temperatures demonstrated a steep narrowing 
of the velocity and density distribution profiles \cite{aem95,bst95,akm97}.
This was explained by the occurrence of the Bose-Einstein condensation (BEC) phenomenon,
earlier known to exist in noble gases, such as helium-4 \cite{lo38}, 
and excitonic semiconductors at low temperatures \cite{swm90}.

From the modern point of view,
Bose-Einstein condensates fall under a wide category of 
quantum Bose liquid.
This special kind of matter includes hydrodynamic media formed not only by integer-spin particles 
but also by the ``bosonized'' combinations of fermions.
The standard theory of quantum Bose liquids is far from its completion,
due to the obvious complexity of this matter and large number of phenomena involved,
as we shall discuss below.   

Historically, 
it was the Gross-Pitaevskii (GP) model, which was first proposed for describing Bose-Einstein condensation \cite{gr61,pi61}.
This is essentially a hydrodynamic version of a quantum
many-body model whose 
condensate particles' interaction potential was truncated at the two-body contact term,
within the perturbation approach's paradigm \cite{psbook,gor20}.
In this model, the cubic nonlinear Schr\"odinger equation naturally occurs as an equation 
for the condensate's wavefunction.
This approach thus assumes that the pair interaction is largest in magnitude when it comes to the Bose-Einstein condensation.
This assumption does seem acceptable for the diluted condensates, such as vapours of alkali atoms,
where interparticle distances 
are sufficiently large to neglect the multi-body (three or more) collisions.
However, as the density grows, multi-body interactions should come into play, 
even for laboratory quantum liquids \cite{ef71,ef73,kms11}, let alone quantum fluids in high-energy physics
and astrophysics.
On top of that, effects of lower-dimensionality and nontrivial vacuum's presence can come into play and somewhat
narrow an applicability range of the Gross-Pitaevskii model.

This inspired further corrections to the Gross-Pitaevskii model, 
\textit{e.g.}, in the form of adding higher-order polynomial terms
to the (formerly) cubic Schr\"odinger equation, to mention just some literature examples \cite{ks92,cr04,crt04}.
Therefore, the question remains whether these polynomial terms are exact,
or they actually represent an infinite series expansion of some non-polynomial function of condensate density.

The conceptually different model of the Bose-Einstein condensate, defined by a wave equation with the
non-polynomial (logarithmic) nonlinearity, was proposed in \cite{z11appb}.
Subsequently,
it was shown that the logarithmic nonlinearity should universally occur in the systems
with the following properties:
(i) they allow a 
hydrodynamic description in terms of the fluid wavefunction (cf. \cite{ry99}),
and (ii) their particles' characteristic interaction potentials are 
substantially larger than kinetic energies \cite{z18zna}. 
A profound relation between logarithmic \schrod~equations and quantum information entropy should be mentioned as well \cite{bra91,az11}.
Fruitful applications of the logarithmic models were found in various 
physical systems \cite{z12eb,dmf03,bgl12,z18epl,cch18,z19ijmpb,sz19,z20un1,z21ltp,lz21cs,z22ijmp,zs22}.
Extensive mathematical studies of logarithmically nonlinear systems were performed as well,
to mention only very recent results \cite{cf21,ff21,jx22,lpa21,la22,kj22mpl,rzz22,cs22dc,wzf22}.

In this paper,
we perform a comparative study of the propagation of sound pulses in cigar-shaped (effectively one-dimensional) Bose-Einstein condensates in Gross-Pitaevskii and logarithmic models.
We shall use the Thomas-Fermi approximation as well as consider a linear regime at some stage. 
This is usually believed to deliver a clear analytical picture of the underlying physics,
while avoiding technical complications.
In the next section we give a brief description of both models.
In section \ref{s:dyn}, we consider their thermodynamical properties,
and in section \ref{s:snd} we study the propagation of small fluctuations of density.
Some discussion and conclusions are presented in section \ref{s:con}.

\scn{The models}{s:mod}
Both Gross-Pitaevskii and logarithmic Bose-Einstein condensates are described by a condensate wavefunction
$\Psi = \Psi (\textbf{x},t)$,
which is normalized to a number $N$ of condensate particles of mass $m$:
\be\lb{e:norm}
\int_\vol |\Psi|^2 \dvol  = 
\int_\vol n \, \dvol = N
,\ee 
where $n = \rho/m = |\Psi|^2$ is particle density 
and $\vol$ is the volume of the condensate.
This wavefunction obeys a minimal $U(1)$-symmetric nonlinear Schr\"odinger equation
\ba
i \hbar \partial_t \Psi
=
\left[-\frac{1}{2} \hbar \cf \, \vena^2
+
V (\textbf{x})
- 
F (|\Psi|^{2})
\right]\Psi
,\label{e:oF}
\ea
where 
$\cf = \hbar/m$,
$V (\textbf{x})$ is an external field potential which usually describes trapping fields and confining vessel's boundaries in BEC experimental setups.
Furthermore, 
$F (n)$ is a preselected function of particle density $n$,
which defines the condensate's type; 
in our case:
\be\lb{e:fcases}
F (n) =
\left\{
\baa{lr}
- U_0 n ,& \ \text{(Gross\textendash Pitaevskii 
~BEC)}\\
\\
b \ln{(n/n_c)} ,& \text{(Logarithmic ~BEC)}
\eaa
\right.
\ee
where $U_0$ is the strength of the two-body interaction,
and
$b$ is the logarithmic nonlinear coupling, and $n_c$ is a critical density value
(at which the logarithmic term is nil).
These
parameters have different physical meaning and origins.
The parameter $U_0$ is a fixed parameter of the GP model, which can be
expressed in terms of atom-atom collisions:
\be   
U_0 = 4 \pi 
\hbar a \cf
,
\ee
where $a$ is the scattering length. 

The parameter $b$ of the logarithmic model is
linearly related to the wave-mechanical temperature $T_\Psi$,
which is defined as a thermodynamical conjugate
of the Everett-Hirschman entropy function
$ 
S_\Psi 
= -\int_\vol |\Psi|^2 \ln{(|\Psi|^2/n_c)} \dvol
$. 
In case of Bose-Einstein condensates, 
this is quantum information entropy which emerges from the logarithmic term
when one averages the logarithmic nonlinear wave equation 
in a Hilbert space
of wavefunctions $\Psi$ \cite{bra91,az11}.
It becomes the thermodynamical entropy of the condensate
$ 
S 
= -\int_\vol n \ln{(n/n_c)} \dvol
$. 
Correspondingly,
it is conjectured that $T_\Psi$
is
linearly related 
to the conventional (thermal) temperature $T$ \cite{z18zna},
thus we can assume here that 
\be
b  \sim T_\Psi \sim T
, 
\ee
or
\be\lb{e:logtemp}
b  = \chi (T - T_c)
,\ee
where $T_c$ is a critical temperature at which the logarithmic term vanishes, 
and 
$\chi$ would be a scale constant dimensionless in units where the Boltzmann constant is one.
This formula indicates that $b$ is not a fixed parameter of the model, but its value
depends on the environment of the logarithmic condensate.

One can see that the attractive/repulsive behaviour for the Gross-Pitaevskii condensate
is fixed by a sign of the strength $U_0$,
while the logarithmic condensate can switch between attractive and repulsive regimes as density
goes across the value $n_c$.
This leads to many profound effects which are only pertinent to the logarithmic Bose liquids, including the
enhanced stability \cite{az11,z21ltp,z17zna}. 

\scn{Thermodynamics and fluid analogy}{s:dyn}
Let us first derive an expression for the chemical potential $\mu$,
which is defined as the Lagrange multiplier 
that ensures constancy of number $N$ under arbitrary variations of wavefunction \cite{psbook}.
Following the standard procedure, we use the stationary ansatz
$\Psi (\textbf{x},t) = \psi (\textbf{x}) \exp{(- i \mu t/\hbar)}$,
to obtain from \eqref{e:oF} the general expression for the chemical potential:
\ba
\mu
=
- 
\frac{1}{2} \hbar \cf \,
\frac{ \vena^2 
\psi}{\psi}
+
V (\textbf{x})
- 
F (|\psi|^{2})
,\label{e:chempot}
\ea
where $F (n)$ would be given by equations \eqref{e:fcases} for the models in question.
For example,
for a free condensate, $V (\textbf{x}) \equiv 0$, 
we obtain
\be\lb{e:cpcases}
\mu \approx
\left\{
\baa{lr}
U_0 |\psi|^2 = U_0 n,\\
\\
- b \ln{(|\psi|^2/n_c)} = - b \ln{(n/n_c)}, 
\eaa
\right.
\ee
where the approximation indicates that we neglected the kinetic energy term, 
trapping fields and boundary effects.
Here and
in equations below, we shall omit the model titles,
assuming that top and bottom lines after a curly bracket
refer to the Gross-Pitaevskii and logarithmic condensates, respectively.

One can see that, similarly to the attractive/repulsive behaviour discussed earlier, 
the chemical potential's sign in the case the Gross-Pitaevskii condensate
is fixed by a sign of the strength $U_0$,
while the logarithmic condensate's $\mu$ switches its sign as density
goes across the value $n_c$.
Thus, adding or removing particles to the logarithmic condensate is energetically favorable
in two of the four regions 
$\{n \gtrless n_c,\ T \gtrless T_c \}$,
and unfavorable in others.
This, of course, has a crucial influence upon a stability of the free logarithmic condensate, as mentioned above.
In presence of an external potential, the picture gets slightly more complicated but the analysis can be easily 
done by analogy.

Furthermore, let us deduce an equation of state from equation \eqref{e:oF}.
To do that, we use the analogy between \schrod-type equations and inviscid isentropic flows.
This analogy has been known for a long time - 
since works by de Broglie and Madelung \cite{bro26,mad27},
and subsequently it was rediscovered few times \cite{halbook,fro66,spi80,ry89}, 
to mention only examples known to the author.
For the logarithmic \schrod~equation specifically, the fluid-\schrod analogy was derived in \Ref \cite{dmf03}, although with some error in the equation of state; 
the correct equation of state and derivation can be found in \Refs \cite{az11,bgl12}. 
For the equation \eqref{e:oF} with a general function $F (n)$,
the fluid-\schrod~analogy was derived in \Ref \cite{z19mat}.

According to the analogy,  
the wave equation \eqref{e:oF} can be rewritten in the hydrodynamic form
by using the Madelung ansatz \cite{bro26,mad27},
from which one can deduce the equation of state and related values.
In the leading order approximation with respect to Planck constant,
we can neglect second derivatives of density (which is robust as long as we do not consider
shock waves and other fluctuations with a non-smooth density profile),
the hydrodynamic equations have the perfect-fluid form,
from which
we obtain the equation of state $p=p(n)$ and speed of sound $c$, respectively, as:
\be\lb{e:cappapp}
p 
\approx
- 
\int\!
n F'(n)\, \drm n
,
\ \
c
\approx
\sqrt{
-
n F'(n)/m
}
,
\ee
where prime denotes an ordinary derivative.
Therefore,
using the definitions \eqref{e:fcases},
we obtain for pressure:
\be\lb{e:pcases}
p 
\approx
\left\{
\baa{lr}
n^2 U_0/2,\\ 
\\
- b n, 
\eaa
\right.
\ee and velocity of sound:
\be\lb{e:ccases}
c 
\approx
\left\{
\baa{lr}
\sqrt{ n U_0/m},\\
\\
\sqrt{- b/m}, 
\eaa
\right.
\ee
where we omitted the approximation sign from now on.
One can see for that the squared velocity of sound is proportional to density in the Gross-Pitaevskii case, 
whereas for the logarithmic fluid it is independent of density. 

A direct comparison of the formulae above indicates that the logarithmic condensate is actually 
closer to the ideal gas 
(whose pressure is a linear function of density)
than the Gross-Pitaevskii condensate is.
Therefore, if one does a Taylor series expansion of the general barotropic fluid's
equation of state with respect to density,
then the first-order term (apart from a constant) would represent  
the logarithmic liquid, whereas the Gross-Pitaevskii term would be only a second-order (quadratic) correction.

\scn{Sound in cigar-shaped condensates}{s:snd}
Let us consider a cloud or lump of BEC whose transverse
dimensions $R$ being significantly less than its axial dimension $H$ aligned along $z$ axis. 
In this setup, we assume that sound pulses have characteristic length scale
$\ell$ in the axial direction, $R \ll \ell \ll H$,
which is usually what takes place in experiments like \cite{aem95,akm97}. 
Along the lines of those experimental setups,
we expect that vertical variations of the confining potential 
are small compared to the length $\ell$,
as well as that the particles density value being
sufficient for the Thomas-Fermi approximation to be
robust.

Since the problem becomes essentially one-dimensional,
the cloud can be characterized by its velocity $v=v(z)$
and linear particle density
\be\lb{e:surf}
\sigma = \sigma (z) = \int n (\textbf{x})\, \drm x \drm y
,
\ee
where $(x,y)$ are coordinates in transverse dimensions.
Because in the previous section we established that the nonlinear \schrod~ equations for both models can be
approximately rewritten in
a hydrodynamic perfect-fluid form, we can derive hydrodynamic equations for $v$ and $\sigma$
by integrating over condensate lump's transverse coordinates \cite{kp97}.
From the continuity and Euler equations,
we obtain, respectively:
\ba
&&
\pDer{\sigma}{t} +
\pDer{(\sigma v) }{z} = 0
,\\&&
m \sigma \Der{v}{t} +
\sigma \pDer{V}{z} = - \int \pDer{p}{z} \, \drm x \drm y 
,
\ea
where the external potential $V$ occurs due to the trap and electromagnetic interactions,
and pressure $p$ is given by either of the formulae \eqref{e:pcases}.

While the continuity equation remains common for both models,
Euler equations must be derived separately.
Using expressions \eqref{e:pcases}, we obtain
\be\lb{e:eucases}
m \sigma \Der{v}{t} +
\sigma \pDer{V}{z} 
=
\left\{
\baa{lr}
\displaystyle
- U_0 \int n \pDer{n}{z} \, \drm x \drm y, \\
\\ \displaystyle
b \int \pDer{n}{z} \, \drm x \drm y, 
\eaa
\right.
\ee
where we also assumed a constant $b$ in the logarithmic case, remembering its thermal interpretation \eqref{e:logtemp}.

In order to simplify these formulae,
let us recall that in 1D case the change of density $n$ can be expressed
as $\drm n = \drm \sigma/\area$, where $\area$ the area of lump's cross section.
Therefore, the Euler equations \eqref{e:eucases} become
\be\lb{e:eucases2}
m \sigma \Der{v}{t} +
\sigma \pDer{V}{z} 
=
\left\{
\baa{lr}
\displaystyle
- \frac{U_0}{\area}  \sigma \pDer{\sigma}{z}  ,\\ 
\\ \displaystyle
b  \pDer{\sigma}{z} , 
\eaa
\right.
\ee
where we also recalled the definition \eqref{e:surf}.

Let us consider the linear approximation, where we also assume that the variation of the external 
potential in the $z$ direction is small compared to other terms.
In case of the Gross-Pitaevskii condensate, this would also mean that one can neglect spatial variations
of the particle density averaged over the lump's cross section.
Altogether, from \eqref{e:eucases} one can derive 
$$ 
\pDer{}{z}
\left(
\sigma
\Der{v}{t}
\right) 
\approx
\left\{
\baa{lr}
\displaystyle
- \frac{U_0}{m}  \pDer{}{z} \left(\frac{\sigma}{\area} \pDer{\sigma}{z}\right) \approx 
- \frac{\bar n U_0}{m} \pDer{^2 \sigma}{z^2}
  ,\\ 
\\ \displaystyle
\frac{b}{m}  \pDer{^2 \sigma}{z^2} , 
\eaa
\right.
$$ 
where $\bar n = \sigma/ \area$ is the particle density averaged over the lump's cross section.
Finally,
taking the total time derivative of the continuity equation and using the last formula,
we obtain
\be\lb{e:wavcases}
\pDer{^2 \sigma}{t^2} 
\approx
\left\{
\baa{lr}
\displaystyle
\frac{\bar n U_0}{m}  \pDer{^2 \sigma}{z^2}  ,\\ 
\\ \displaystyle
- \frac{b}{m}  \pDer{^2 \sigma}{z^2}   , 
\eaa
\right.
\ee
where we are left only with terms which are linear with respect to $\sigma$.

If we assume that $U_0$ is positive,  and $b$ is negative, $b = - |b|$ (\textit{i.e.}, $\chi >0$ and $T < T_c$),
then we obtain the hyperbolic-type partial differential equation in both cases.
This indicates that small density fluctuations propagate in
the direction perpendicular to the lump's cross section
with the velocities
\be\lb{e:ccases2}
\bar c
=
\left\{
\baa{lr}
\displaystyle
\sqrt{\bar n U_0/m}    ,\\ 
 \\ \displaystyle
\sqrt{|b|/m}   , 
\eaa
\right.
\ee
which 
are similar to the corresponding expressions for the homogeneous gas, 
except for the Gross-Pitaevskii condensate 
density must be replaced by the average density over the cross section of
the lump.
It also
reaffirms the one-dimensional character of sound propagation in Bose-Einstein condensates of both types.

\scn{Conclusion}{s:con}
In this paper, we performed a comparative study of the propagation of 
small fluctuations of density in cigar-shaped Bose-Einstein condensates in Gross-Pitaevskii and logarithmic models, using the Thomas-Fermi and linear approximations.

It is shown that the propagation of sound pulses  is effectively one-dimensional in both cases.
The difference between them is that speed of sound scales as square root of particle density in a case
of the Gross-Pitaevskii condensate,
whereas this velocity is constant in a case of the logarithmic condensate.
It should be emphasized that these results were obtained assuming a number of approximations,
which do not necessarily hold in a presence of large density inhomogeneities.
For example, logarithmic condensate, being essentially nonlinear, is known to form such inhomogeneities,
because it behaves more like quantum liquid than gas \cite{az11,z12eb,z19ijmpb}.
Apart from the nonlinear phenomena, shock waves and other objects with effectively non-smooth density profiles
can violate the above-mentioned approximations too.

Study of nonlinear and shock-wave effects in one-dimensional quantum Bose liquids under more general conditions goes beyond a scope of this report,
and thus it will be a subject of future work.

\funding{This work is based on the research supported by the Department of Higher Education and Training of South Africa
and in part by the National Research Foundation of South Africa (Grants Nos. 95965, 131604 and 132202).}


\begin{acknowledgments}
Author gratefully acknowledges an invitation and a full waiver on this article's processing charges by Multidisciplinary Digital Publishing Institute to get it published in the Open Access form. 
This work is based on the research supported by the Department of Higher Education and Training of South Africa
and in part by the National Research Foundation of South Africa (Grants Nos. 95965, 131604 and 132202).
\end{acknowledgments} 


\conflictsofinterest{The author declares no conflict of interest.}





\reftitle{References}

\end{document}